\documentclass[onecolumn]{aa}
\usepackage{graphicx}
\usepackage{txfonts}
\usepackage{psfig}
\usepackage{epsfig}

\def \hcm {\hbox {\ifmmode $ cm$^{-2}\else cm$^{-2}$\fi}}

\def\approxgt{\mathrel{\hbox{\rlap{\lower.55ex \hbox {$\sim$}}
        \kern-.3em \raise.4ex \hbox{$>$}}}}
\def\approxlt{\mathrel{\hbox{\rlap{\lower.55ex \hbox {$\sim$}}
        \kern-.3em \raise.4ex \hbox{$<$}}}}

\begin{document}

\title{The optical--near-infrared colour of the host galaxies of 
BL Lacertae objects}

\author{J.K. Kotilainen\inst{1} \and R. Falomo\inst{2}}

\offprints{J.K. Kotilainen}

\institute{Tuorla Observatory, University of Turku, V\"ais\"al\"antie 20, 
FIN--21500 Piikki\"o, Finland\\
\email{jarkot@utu.fi}
\and
INAF -- Osservatorio Astronomico di Padova, Vicolo dell'Osservatorio 5, 
35122 Padova, Italy\\
\email{falomo@pd.astro.it}
}

\date{Received; accepted }

\abstract{
We present $H$-band (1.65 $\mu$m) imaging of 23 low redshift (z $\leq$ 0.3) 
BL Lac objects that were previously investigated by us in the 
optical $R$-band. For all the observed objects, the host galaxy is clearly 
detected and well represented by an elliptical model, with average 
scale length R(e) = 7.2$\pm$3.6 kpc and average absolute magnitude 
M(H) = -25.9$\pm$0.6. BL Lac hosts are therefore luminous (massive) 
elliptical galaxies indistinguishable from those of low redshift 
radio galaxies and inactive ellipticals. The best-fit $H$-band 
Kormendy relation of $\mu$$_e$ = 3.7 log r$_e$ (kpc) + 14.8 mag arcsec$^{-2}$ 
is in agreement with that obtained for normal ellipticals and radio galaxies. 
This structural and dynamical similarity indicates that all massive 
elliptical galaxies can experience nuclear activity without 
significant perturbation of their global structure. 

The new observations are combined with literature data 
(Kotilainen et al. 1998; Scarpa et al. 2000b; Cheung et al. 2003) of 
18 objects in order to construct a sizeable dataset of 41 BL Lacs with 
determined host characteristics in both optical and NIR. This extended 
sample is used to investigate the optical--NIR colour and colour gradient 
properties of the host galaxies of BL Lacs and to perform a direct comparison 
with other elliptical galaxies with and without nuclear activity. 

The integrated optical/near--infrared colour ($R$--$H$ = 2.2$\pm$0.4) and 
colour gradient ($\Delta$($R$--$H$)/$\Delta$(log r) = -0.38$\pm$0.60) 
distributions of the BL Lac hosts are much wider than those for 
normal ellipticals with old stellar populations, and many BL Lacs have 
bluer hosts and/or steeper colour gradients than those in normal ellipticals. 
The blue colours are most likely caused by a young stellar population, 
and indicates a link between star formation caused by an 
interaction/merging event and the onset of the nuclear activity. 
However, the lack of obvious signs of interaction may require   
a significant time delay between the event with associated SF episodes and 
the start of the nuclear activity.

\keywords{BL Lacertae objects:general -- Galaxies:active -- 
Galaxies: elliptical and lenticular. cD -- Galaxies:nuclei -- 
Galaxies:photometry -- Infrared:galaxies}
}

\titlerunning{NIR imaging of BL Lac hosts}
\authorrunning{Kotilainen \& Falomo}

\maketitle

\section{Introduction}

BL Lac objects are active galactic nuclei (AGN) characterized by luminous, 
rapidly variable UV--to--near-infrared (NIR) non--thermal continuum emission 
and polarization, strong compact flat spectrum radio emission and 
superluminal motion. Similar properties are observed also in flat spectrum 
radio quasars (FSRQ) and the two types of AGN are often grouped together in 
the blazars class. The clearest difference between BL Lacs and FSRQs is that 
the latter have strong broad emission lines of similar strength to 
normal quasars, while in BL Lacs they are very weak or absent 
(e.g. Scarpa \& Falomo 1997). 

These extreme properties are commonly interpreted assuming that the observed 
emission is dominated by a relativistically beamed synchrotron jet oriented 
close to our line--of--sight (Blandford \& Rees 1978; Urry \& Padovani 1995). 
This model implies the existence of a more numerous  population of 
intrinsically identical objects but with their jet oriented at larger angles 
to our line--of--sight. In the current unified models for radio--loud AGN 
(e.g. Urry \& Padovani 1995), BL Lacs are low luminosity core--dominated 
(FR I) radio galaxies (RG) with jets pointing close to our line-of-sight, 
while FSRQs are high luminosity lobe--dominated (FR II) RGs. As a consequence 
of this model, any orientation--independent properties of the BL Lac objects 
(e.g. extended radio emission, host galaxies and environments) should be 
identical to those of the parent population (i.e. FR I RGs). 

Optical imaging studies of BL Lacs secured either with groundbased telescopes 
(e.g. Abraham, McHardy \& Crawford 1991; Stickel, Fried \& K\"uhr 1993; 
Falomo 1996; Wurtz, Stocke \& Yee 1996; Falomo \& Kotilainen 1999, 
Pursimo et al. 2002; Nilsson et al. 2003) or with the HST 
(e.g. Falomo et al. 1997b; Jannuzi, Yanny \& Impey 1997; Scarpa et al. 2000a; 
Urry et al. 2000) have shown that virtually all nearby (z $<$ 0.5) objects are 
hosted in giant ellipticals with, in some cases, complex morphology 
(close companions, jets). They have similar magnitude to both FR I and 
FR II RGs (Govoni et al. 2000). Although some claims of a significant disc 
component have been made in some sources (Stocke, Wurtz \& Perlman 1995; 
Wurtz et al. 1996), the HST imaging survey of BL Lacs found overwhelming 
preference for elliptical hosts (Scarpa et al. 2000a; Falomo et al. 2000). 
Taken together, these works have demonstrated that the hosts of (virtually) 
all BL Lacs are luminous (mostly unperturbed) elliptical galaxies located in 
poor environments. Both the global (Urry et al. 2000) and the detailed 
(e.g. ellipticity, isophote twisting and shape) characteristics 
(Falomo et al. 2000) of the host galaxies are indistinguishable from those of 
inactive massive ellipticals.

On the other hand, because most of these studies have been performed in 
only one (usually optical) band, little colour information of BL Lacs exists. 
NIR imaging of small ($<$ 10 sources) BL Lac samples were performed by 
Kotilainen, Falomo \& Scarpa (1998; hereafter K98), Scarpa et al. (2000b; 
hereafter S00) and Cheung et al. (2003; hereafter C03). These studies 
suggested that the optical-NIR colour and the colour gradient of BL Lac hosts 
is similar to those of inactive ellipticals 
(e.g. Peletier, Valentijn \& Jameson 1990) and low redshift RGs 
(e.g. de Vries et al. 1998).

The existing dataset therefore suggests that the nuclear activity does not 
strongly affect the properties of BL Lac hosts, which can be explained by 
the bulk of star formation (SF) occurring at high redshift, followed by 
passive stellar evolution (Stanford, Eisenhardt \& Dickinson 1998). 
Indications of a very old, coeval stellar population have been found for RGs 
and inactive ellipticals (e.g. McCarthy 1993). For example, 
galaxy interactions and mergers, often invoked to explain AGN phenomena, 
do not seem to induce massive SF in BL Lac hosts, possibly due to low 
gas density and/or different timescales of SF and nuclear activity. 
This result has obvious implications for the host galaxy-AGN connection and 
the evolution of AGN. For example, since many inactive ellipticals appear to 
have supermassive black holes (e.g. van der Marel 1999), it is possible that 
all ellipticals have the potential to become active with increased accretion. 
 
The colour properties of Bl Lac hosts are, however, based on 
optical/NIR imaging of only 20 objects that were investigated by 
different groups 
(seven sources in K98, ten in S00, eight in C03; five common objects among 
the samples), and need to be confirmed using a larger homogeneous sample. 
In this paper, we present deep high spatial resolution NIR $H$--band 
(1.65 $\mu$m) images of 23 low redshift (z $<$ 0.3) BL Lac objects 
(19 high-frequency-peaked and 4 low-frequency-peaked objects) aimed at 
deriving the NIR properties of their host galaxies. There are two objects 
in common with the previous studies (MS 0158.5+0019 by C03, 
and PKS 2254+074  by K98).

The new observations are combined with previous data in order to form 
a sizeable sample of 41 low redshift BL Lac objects with determined NIR and 
optical host galaxy properties. For all the observed objects we were able 
to study the host galaxy in the optical ($R$-band) using NOT 
(Falomo \& Kotilainen 1999) and/or HST (Urry et al. 2000). This enables us 
to homogeneously investigate the integrated $R$--$H$ colour and its gradient 
for the BL Lacs and to compare them with those of RGs and 
inactive ellipticals. We emphasize that still few colour gradient 
measurements exist for any class of AGN, and measuring them in BL Lacs 
will find immediate implications to studies of other types of AGN. 

In section 2, we describe the observations, data reduction and the method of 
the analysis. Our results are presented in section 3 and conclusions in 
section 4. Throughout this paper, H$_{0}$ = 50 km s Mpc and 
$\Omega_{0}$ = 0 are used. 

\section{Observations, data reduction and modeling of the luminosity profiles}

The observations were carried out at the 2.5m Nordic Optical Telescope (NOT) 
on four nights in July 2002, using the 1024 x 1024 px NOTCam NIR camera with 
pixel scale 0.235$''$ px$^{-1}$, giving a field of view of 4x4 arcmin$^2$. 
The $H$-band (1.65 $\mu$m), corresponding to the minimum in the nucleus/host 
luminosity ratio for low z objects, was used for all the observations. 
The seeing during the observations was excellent, ranging from 0\farcs5 to 
0\farcs9 FWHM (average and median = 0\farcs7). A journal of the  observations 
is given in Table~\ref{sample}. For the majority of the objects the images 
were acquired keeping the target in the field by dithering it across the array 
and acquiring several short exposures at each position. Individual exposures 
were then coadded to achieve the final integration time 
(see Table~\ref{sample}). Only for the most nearby targets, object frames 
were interspersed with sky frames in order to perform an accurate 
sky subtraction. 

Data reduction was performed using the NOAO Image Reduction and 
Analysis Facility (IRAF\footnote{IRAF is distributed by the 
National Optical Astronomy Observatories, which are operated by the 
Association of Universities for Research in Astronomy, Inc., under cooperative 
agreement with the National Science Foundation.}). Bad pixels were identified 
via a mask made from the ratio of two sky flats with different 
illumination level, and were substituted by interpolating across 
neighboring pixels. Sky subtraction was obtained using a median averaged 
sky frame (for nearby sources) or a median averaged frame of the temporally 
closest frames (for most sources). Flat-fielding was made using twilight 
sky frames, and images of the same target were registered and combined using 
field stars as reference points to obtain the final reduced co-added image. 

Observations of standard stars taken from Hunt et al. (1998), were used for 
the photometric calibration, for which we estimate an accuracy of 
$\leq$ 0.1 mag. No K--correction was applied to the host galaxy magnitudes 
as the size of this correction is insignificant at z $\leq$ 0.3 in 
the $H$-band ($\Delta$m(H) $<$ 0.02; Neugebauer et al. (1985). 
No K--correction was applied to the nuclear component, assumed to have 
a power-law spectrum ($f_\nu \propto \nu^{-\alpha}$) with $\alpha$ $\sim$1. 
The interstellar extinction corrections were computed using R-band 
extinction coefficient from Urry et al. (2000) and A$_H$/A$_R$ = 0.234 
(Cardelli, Clayton \& Mathis 1989).

In order to characterize the properties of the host galaxies, 
azimuthally averaged one-dimensional radial luminosity profiles were extracted 
for each BL Lac object and for various stars in the frames down to 
surface brightness of $\mu$(H) $\sim$23 mag arcsec$^{-2}$. 
As all the sample objects are well resolved and almost all are known 
from optical studies to have quite round and undisturbed host galaxies 
(Falomo \& Kotilainen 1999; Urry et al. 2000), we are confident that 
the 1D modelling is not affected by unusual features in the host galaxy. 
Note that the most problematic case in this respect is 1ES 1440+122, 
which has a companion galaxy at the redshift of the BL Lac. They are 
undergoing weak interaction that is likely responsible for the slightly 
boxy isophotes of 1ES 1440+122 (Falomo et al. 2000) and the slight 
isophotal twisting of the companion (Heidt et al. 1999). There is, however, 
no evidence for disturbed morphology of 1ES 1440+122 in the 2D HST analysis 
(Falomo et al. 2000) and therefore we assume that the BL Lac host is 
unperturbed even in this case.

The luminosity profiles were decomposed into the point source and 
galaxy components by an iterative least-squares fit to the observed profile. 
We fit the data  using both elliptical $r^{1/4}$ (de Vaucouleurs law) and 
exponential disc models to represent the host galaxy. 
However, and consistently with the results from previous optical and 
NIR studies, in no case did the disc model give a better fit than 
the elliptical one. We estimate the uncertainty of the derived host galaxy 
magnitudes to be $\sim$$\pm$0.2 mag. 

\section{Results and discussion}

In Fig.~\ref{blfig1} we show the $H$-band contour plots of all the observed 
BL Lac objects. We are able to clearly detect the host galaxy in all the 
observed targets. In Fig.~\ref{blfig2}, we show the azimuthally averaged 
$H$-band radial luminosity profiles of each BL Lac object, together with 
the best--fit models overlaid. The results of the best-fit modelling are 
summarized in Table~\ref{hostprop}. 

Three host galaxies in the sample were previously studied in the NIR 
(MS 0158.5+0019 studied by C03, and PKS 2254+074 studied by K98; 
MS 0317.0+1834 by Wright et al. 1998). The comparison between these studies is 
presented in Table~\ref{compbl}. 
For MS 0158.5+0019, the derived host magnitudes are in good agreement,
whereas we find larger bulge scale length.
C03 find much redder optical-NIR colour 
and much steeper colour gradient 
than in this work.
PKS 2254+074, already studied by K98, was re-observed here because K98 
obtained only a short exposure (10 minutes) under poor sky conditions and 
the object was situated in a bad area of the array during their observations. 
Consequently, the host properties were less accurately than for the other 
resolved BL Lacs in K98. With improved data quality, we detect a much brighter 
host galaxy 
with a more reasonable bulge scale length, 
slightly redder colour 
and steeper colour gradient.
Finally, MS 0317.0+1834 was studied with poorer quality data in the $K$-band 
by Wright et al. (1998), who marginally resolved 
somewhat fainter and a factor of two larger host than what is found in 
this work.

\subsection{Morphology and luminosity of the host galaxies}

In Fig.~\ref{mhz}, we show the $H$--z (Hubble) diagram for the BL Lac hosts 
(this work, K98, S00 and C03), compared with  the established relation for RGs 
(solid line; Willott et al. 2003). The BL Lac hosts follow the same trend 
in the $H$--z diagram as that traced by RGs, although there is a tendency 
for the BL Lac hosts to be on average fainter than the RG sample. The same 
trend is visible in Fig.~\ref{Mhz}, where we report the $H$--band host galaxy 
absolute magnitude vs. redshift for the BL Lacs (this work, K98, S00 and C03) 
and RGs. 
In the NIR, most of the host galaxies of low redshift BL Lacs are encompassed 
within M* and M*-2 (where an M$^*$ galaxy has M(H) = --25.0$\pm$0.3; 
Mobasher, Sharples \& Ellis 1993), in agreement with previous results derived 
in the optical (e.g. Falomo \& Kotilainen 1999; Urry et al. 2000).

The comparison among the various samples of the average NIR 
absolute magnitudes of the host galaxies is given in Table~\ref{avgprop}.  
These samples span a moderately large range in redshift from z$\sim$0.03 up to 
z$\sim$0.3. All magnitudes were extinction-corrected and transformed into our 
adopted cosmology and into the $H$-band, assuming the average colour of 
giant ellipticals, $H$--$K$ = 0.22 (Recillas-Cruz et al. 1990). The average 
absolute magnitude of the low redshift (z$\leq$0.3) BL Lac hosts in this work 
is M(H) = --25.8$\pm$0.7, indistinguishable from that of the other 
BL Lac samples. For the total combined sample of 41 BL Lacs, we obtain 
average M(H) = --25.9$\pm$0.6. 
As found in the optical by e.g. Wurtz et al. (1996), 
Falomo \& Kotilainen (1999) and Urry et al. (2000), BL Lac hosts have 
slightly lower luminosities in the NIR than nearby brightest cluster member 
galaxies (BCM; M(H) = --26.3$\pm$0.3; Thuan \& Puschell 1989) and the 
(mainly 3C) RGs at z $<$ 0.3 studied by  Willott et al. (2003).

The average bulge scalelength of the 23 low redshift BL Lac hosts in this work 
is R(e) = 7.8$\pm$4.1 kpc, (R(e) = 7.2$\pm$3.6 kpc. for the combined sample 
of 41 BL Lacs). Slightly smaller (K98 and C03) and larger (S00) values 
were found in previous studies (see Table~\ref{avgprop} for details). 
Consistently with optical studies, therefore, the BL Lac host galaxies are 
with a few exceptions larger than normal ellipticals 
(e.g. Capaccioli, Caon \& D'Onofrio 1992), but smaller than low redshift RGs 
studied in the optical by Govoni et al. (2000; R(e) = 16$\pm$10 kpc) and in 
NIR by Taylor et al. (1996; R(e) = 26$\pm$16 kpc).

\subsection{Surface brightness - effective radius (Kormendy) relation}

NIR observations map the bulk of the slowly evolving stellar populations 
of galaxies, and are therefore ideal for investigating luminosity and 
galaxy morphology. We have investigated the $H$-band Kormendy relation, 
which is a projection of the fundamental plane, relating the effective radius, 
$r_e$, to the surface brightness at that radius, $\mu_e$ 
(Kormendy 1977; Djorgovski \& Davis 1987; Kormendy \& Djorgovski 1989). 
This relation is closely related to the morphological and dynamical 
structure of galaxies, and to their formation processes. 

We combined our $\mu_{e}$ and $r_{e}$ data with those from K98, S00 and C03 
to construct the Kormendy relation in the NIR for the largest available 
sample of BL Lacs (see Fig \ref{muere}). The effective surface brightnesses 
of the hosts have been corrected for galactic extinction and 
cosmological dimming (10 $\times$ log(1+z)). The $K$-band data from C03 
were converted to the $H$-band assuming $H-K=0.22$ 
(Recillas-Cruz et al. 1990). We find a best-fit linear relation for the 
BL Lacs to be 
$\mu_{e}$(H) = 3.7 ($\pm$0.3) log $r_{e}$ + 14.8 ($\pm$0.3) mag arcsec$^{-2}$. 
This is very similar to that derived for normal elliptical galaxies 
($\mu_{e}$(H) = 4.3 log $r_{e}$ + 14.3 mag arcsec$^{-2}$; 
Pahre, Djorgovski \& de Carvalho 1995). 

This similarity between BL Lacs and normal ellipticals (which also holds for RGs, 
e,g. Govoni et al. 2000) suggests that BL Lac hosts (and RGs) are 
dynamically very similar and occupy the same region of this projection of 
the fundamental plane as normal ellipticals. This  indicates that 
the formation processes and the structure of galaxies hosting radio sources 
are similar to those of radio-quiet ellipticals. This result is reinforced 
by spectroscopic studies of the fundamental plane (Bettoni et al. 2001; 
Barth, Ho \& Sargent 2003; Falomo, Kotilainen \& Treves 2002; 
Falomo et al. 2003) and supports the picture in which all massive 
elliptical galaxies have the potential to experience a phase of 
nuclear activity with little influence on their global structure. 

\subsection{Host galaxy colours}

The host galaxies of all the low redshift BL Lacs in our sample were 
previously studied by us in the optical (Falomo et al. 1997a,b; 
Falomo \& Kotilainen 1999; Urry et al. 2000). The new observations for 
21 objects represent the first study of them performed in the NIR. These data, 
combined with other NIR and optical data, allow us to address  the issue 
of the optical--NIR colour of the BL Lac host galaxies using a sizeable 
and homogeneous dataset. The integrated rest-frame colours of the 
host galaxies are given in Table~\ref{colour}. The average and median 
integrated $R$--$H$ host colour of the BL Lac sample in this work is 
$R$--$H$ = 2.2$\pm$0.4 and $R$--$H$ = 2.3, and for the total combined sample 
of 41 BL Lacs $R$--$H$ = 2.3$\pm$0.4 and $R$--$H$ = 2.4. 

It is well known that the integrated colours of elliptical galaxies become 
bluer towards fainter magnitudes (e.g. Kodama \& Arimoto 1997 and 
references therein). Observations of large samples of cluster galaxies 
(e.g. Bower et al. 1992a,b) have also shown that this colour-magnitude 
(C-M) relation is virtually identical for different clusters. This relation 
is likely connected with the processes of galaxy formation and it may 
depend on the combined effects of age and metallicity of the underlying 
dominant stellar populations.

According to this relation, therefore, the comparison of the colour properties 
of BL Lac hosts and other types of ellipticals must take into account 
the dependence of the colour on the luminosity of the galaxy. 
In Fig.~\ref{hrh} we report the optical-NIR C-M diagram of the full sample of 
BL Lac host galaxies, compared with those of elliptical and S0 galaxies in 
the Virgo and Coma clusters (Bower et al. 1992a,b). It is immediately 
clear that BL Lac hosts do not follow the C-M relation of ellipticals. 
At variance with normal ellipticals, the BL Lac hosts exhibit a much 
broader range of colours, and their colours 
appear to be systematically bluer than those of the corresponding 
(inactive) ellipticals. Especially, note the very blue tail with R-H $<$ 2.0 
exhibited by eight BL Lacs in the sample. This behavior is similar to that 
found for low redshift RGs by Govoni et al. (2000; see their Fig. 12). 

A possible explanation for this trend (as also suggested by 
Govoni et al. 2000) is that the C-M relation for elliptical galaxies 
breaks down at the highest luminosities. Unfortunately, to our knowledge 
there are no published optical and NIR observations of normal inactive 
galaxies in the luminosity range covered by the BL Lacs and therefore 
the comparison must be based on the extrapolation of the behavior of 
the C-M relation. Indeed, the BL Lac hosts (and RGs in Govoni et al.) 
cover preferentially the bright end of the luminosity function 
of ellipticals, whereas very few of the ellipticals in Fig.~\ref{hrh} 
are equally luminous (but those few exhibit markedly redder colour than 
the active ellipticals). 

On the other hand, assuming that the C-M relation derived for lower luminosity 
galaxies extends towards higher luminosity, the obvious conclusion for the 
blue colours is that the BL Lac hosts exhibit a signature of recent SF. 
The broad range of colours could then reflect differences between objects 
in the epoch of the last burst of SF. Objects with very blue colours would 
have experienced the most recent SF episode, whereas those with red colours 
would be consistent with a single old stellar population. In this view, 
therefore, the bluer colours and the broader colour distribution of 
BL Lac hosts and low redshift RGs could mark the signature of the connection 
between the event(s) that have triggered and/or fuelled the nuclear activity 
and triggered a young stellar population.

This scenario is corroborated by the results on the colour of low redshift 
(z $<$ 0.2) quasar hosts obtained by Jahnke, Kuhlbrodt \& Wisotzki (2003) 
(Fig.~\ref{hrh}). For the 10 quasars hosted in elliptical galaxies they found 
systematically bluer colours ($R$-$H$ = 1.8$\pm$0.3) with respect to those 
of inactive ellipticals, and from population synthesis modelling found 
for them evidence for a few Gyr stellar population, much younger than 
expected in old evolved elipticals. Similarly blue host colours were found for 
low redshift, intermediate luminosity AGN by Schade, Boyle \& Letawsky (2000). 

\subsection{Colour gradients}

It is well known that nearby inactive ellipticals have negative 
colour gradients in the sense that the galaxies become bluer with 
increasing radius (e.g. Peletier et al. 1990). This observed colour gradient 
has been ascribed to the effect of either age or metallicity gradients 
(e.g. Wise \& Silva 1996). Both effects can produce a colour gradient, 
although the latter appears to agree better with the observed gradients in 
nearby elliptical galaxies (Tamura et al. 2000). Independently of 
the origin of the colour gradient, it is important to compare the 
colour gradient of inactive ellipticals with that of active galaxies. 

For our sample of observed BL Lac objects, we computed the radial 
($R$--$H$) colour profiles (see Fig.~\ref{blfig3}), using the host galaxy 
luminosity profiles from ground-based $R$-band images 
(Falomo \& Kotilainen 1999; Falomo \& Kotilainen, in prep.) when available, 
and from HST images (Scarpa et al. 2000a) for the others. The colour profiles 
against logarithmic radius are usually smooth and well represented by 
a linear fit. Most of the colour profiles (see Table~\ref{colour}) show 
a negative colour gradient 
(average $\Delta$($R$--$H$)/$\Delta$(log r) = -0.38$\pm$0.60). This trend 
with similar amplitude was previously suggested by K98 and S00, but is 
much flatter than that of C03. 
In the full sample of 41 BL Lacs, the average 
$\Delta$($R$--$H$)/$\Delta$(log r) = -0.51$\pm$0.65. However, the gradients in 
the $K$-band study of C03 ($\Delta(R-K)/\Delta log~r = -1.27 \pm 0.73$) may be 
influenced by the use of small NIR effective radii (average 4.2 kpc), 
compared with $\sim$10 kpc in the $R$-band (Falomo \& Kotilainen 1999; 
Urry et al. 2000). Therefore, excluding the data of C03, the average 
colour gradient in the combined sample becomes 
$\Delta$($R$--$H$)/$\Delta$(log r) = -0.34$\pm$0.50.

The amplitude of the negative colour gradient in BL Lacs is  steeper 
than that exhibited by normal inactive ellipticals 
(e.g. $\Delta$($V$--$K$)/$\Delta$(log r) = --0.16$\pm$0.18; 
Peletier et al. 1990; $\Delta$($V$--$K$)/$\Delta$(log r) = --0.26$\pm$0.15; 
Schombert et al. 1993) although the spread of values is quite large.
For nearby RGs, Govoni et al. (2000) also found slightly steeper 
colour gradient ($\langle \Delta(B-R)$/$\Delta log~r\rangle$ = -0.16$\pm$0.17) 
than that of normal ellipticals 
($\langle \Delta(B-R)$/$\Delta log~r\rangle$ = -0.09$\pm$0.07; 
Peletier et al. 1990) 

In Fig.~\ref{colgradhist} we show the distribution  of the R--H 
colour gradient for the combined sample of BL Lacs. For the majority of 
the objects the colour gradient is confined between -1.2 to 0.4 
(average $\Delta$($R$--$H$)/$\Delta$(log r) $\sim$ -0.3) while for 
normal ellipticals (Peletier et al. 1990; Schombert et al. 1993), there is 
a narrower spread. A similar larger spread for the B-R colour gradient 
was found by Govoni et al. (2000) for low z RGs.

Note that there are some exceptions with a slightly positive gradient, 
i.e. bluer colour toward the center, e.g. in Mrk 180, 1ES 1426+428 
and 1ES 1853+671. More intriguingly, we confirm the result of C03 of 
a significant tail in the distribution with very strong red gradients 
toward the center, e.g. in Mrk 501, BL Lac and PKS 2201+04. These colour 
differences could be either intrinsic or due to central dust extinction in 
the host galaxies. Significant amount of dust in the central regions of RGs 
is often detected from HST imaging (e.g. Verdoes Kleijn et al. 1999; 
Capetti et al. 2000) as dust lanes or dusty disks. These small scale features 
are in general, however, undetectable in ground-based observations. 

\section{Summary and Conclusions}

We have presented homogeneous NIR imaging of a sample of 23 low redshift 
(z $\leq$0.3) BL Lac objects that were previously investigated in the optical 
band. All the new observed BL Lacs are clearly resolved, and their 
host galaxies well described in the NIR by an elliptical model. These new 
observations are combined with previous NIR data (K98, S00, C03) in order 
to study the NIR properties of the host galaxies in a sizeable sample 
(41 BL Lacs).

The global properties of the BL Lac host galaxies (average luminosity 
M(H) = --25.8$\pm$0.7 and bulge scale length R(e) = 7.8$\pm$4.1 kpc) 
confirm previous suggestions that also in the NIR BL Lac hosts are luminous 
(massive) early type galaxies with structural and photometrical 
characteristics indistinguishable from those of low redshift RGs and 
inactive ellipticals. The NIR $r_e$ - $\mu_e$ relation for BL Lac 
host galaxies is similar to that of normal ellipticals and RGs. Thus they 
appear to be also 
dynamically similar and occupy the same region of this projection 
of the fundamental plane as other ellipticals. This indicates that all massive 
elliptical galaxies can experience nuclear activity with little influence 
on their global structure. 

There is increasing evidence that the photometric properties of 
elliptical galaxies (at least of those in nearby clusters) can be explained 
in terms of a single burst of SF at high redshift, followed by passive 
stellar evolution (e.g. Stanford et al. 1998 and references therein). 
On the other hand, on-going galaxy interactions and/or 
minor mergers, often invoked to explain and sustain the nuclear activity 
(e.g. Heckman 1990), could induce a recent episode of SF.  

Our NIR study combined with optical data was able to address the issue of 
the optical--NIR colour and colour gradient of BL Lac host galaxies using 
a large and homogeneous dataset. The results confirm and extend 
previous findings (also reported for RGs and quasars) that the R-H colour 
and colour gradient distributions of 
the BL Lac hosts are much wider than those for normal ellipticals with old 
stellar populations, and many BL Lacs have bluer hosts and/or steeper 
colour gradients than those observed in normal ellipticals. Assuming that 
the blue colours are caused by a young stellar population component, 
it may represent the link between the SF episode caused by 
an interaction/merging event and the onset of the nuclear activity. 
However, both the optical and NIR images of BL Lac hosts show few 
obvious signs of interaction (e.g. tidal tails, asymmetries, 
secondary nuclei). This may require  a significant time delay 
(at least 100 Myr) between the event with associated SF episodes and 
the start of the nuclear activity.

\begin{acknowledgements}
Nordic Optical Telescope is operated on the island of La Palma jointly by 
Denmark, Finland, Iceland, Norway and Sweden, in the Spanish Observatorio del 
Roque de los Muchachos of the Instituto de Astrofisica de Canarias. 
This research has made use of the NASA/IPAC Extragalactic Database 
(NED), which is operated by the Jet Propulsion Laboratory, California 
Institute of Technology, under contract with the National Aeronautics and 
Space Administration. JKK acknowledges financial support from the 
Academy of Finland, project 8201017.
\end{acknowledgements}

\begin{figure*}
\centering
\includegraphics[width=15cm]{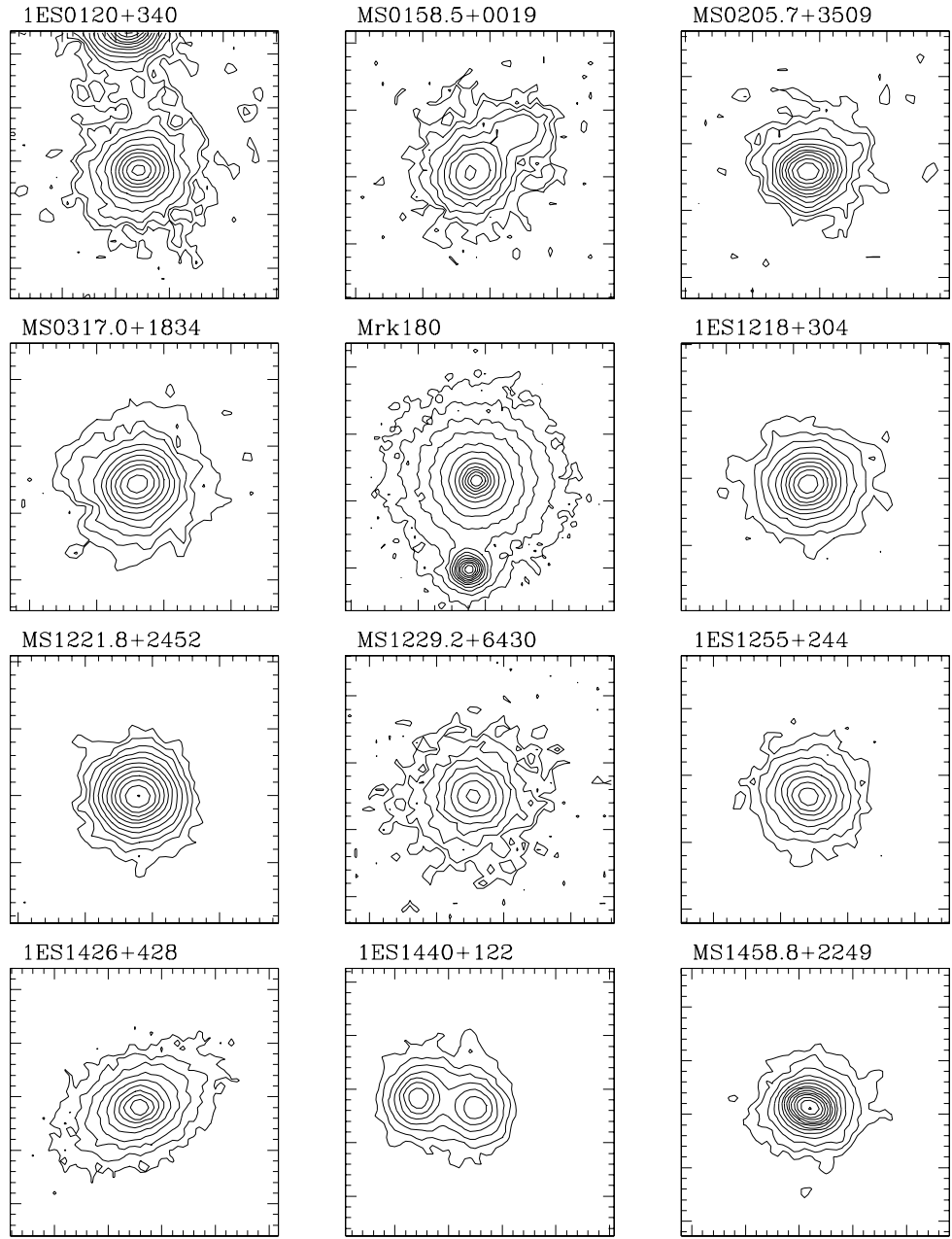}
\caption{
Contour images of the BL Lac objects in the $H$-band. The distance between 
major tick marks is 10 px (2\farcs35).
Successive isophotes are separated by 0.5 mag intervals.
North is up and east to the left. The host galaxy is resolved in all objects. 
}
\label{blfig1}
\end{figure*}
\addtocounter{figure}{-1}%

\begin{figure*}
\centering
\includegraphics[width=15cm]{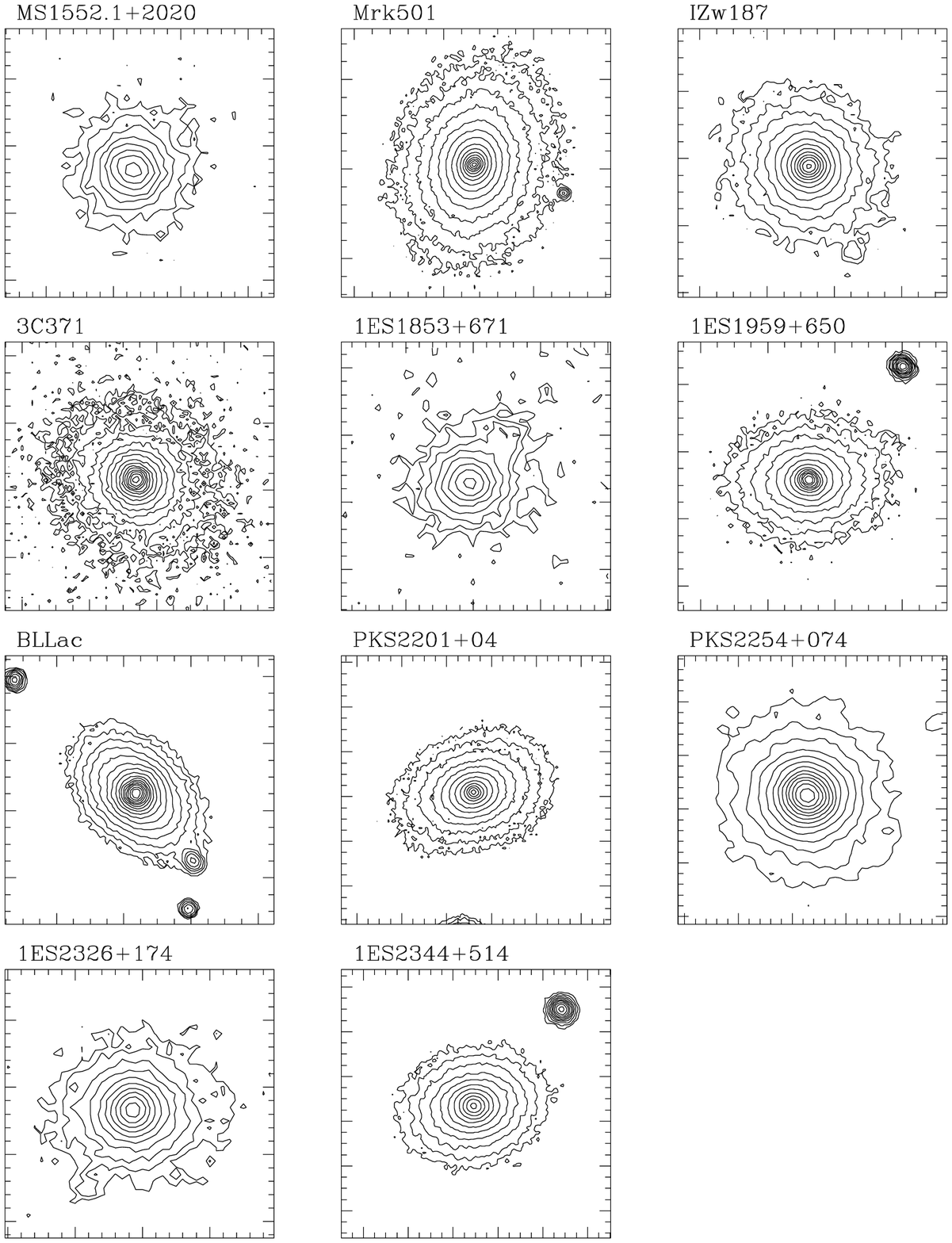}
\caption{Continued.
}
\label{blfig1}
\end{figure*}

\begin{figure*}
\centering
\includegraphics[width=15cm]{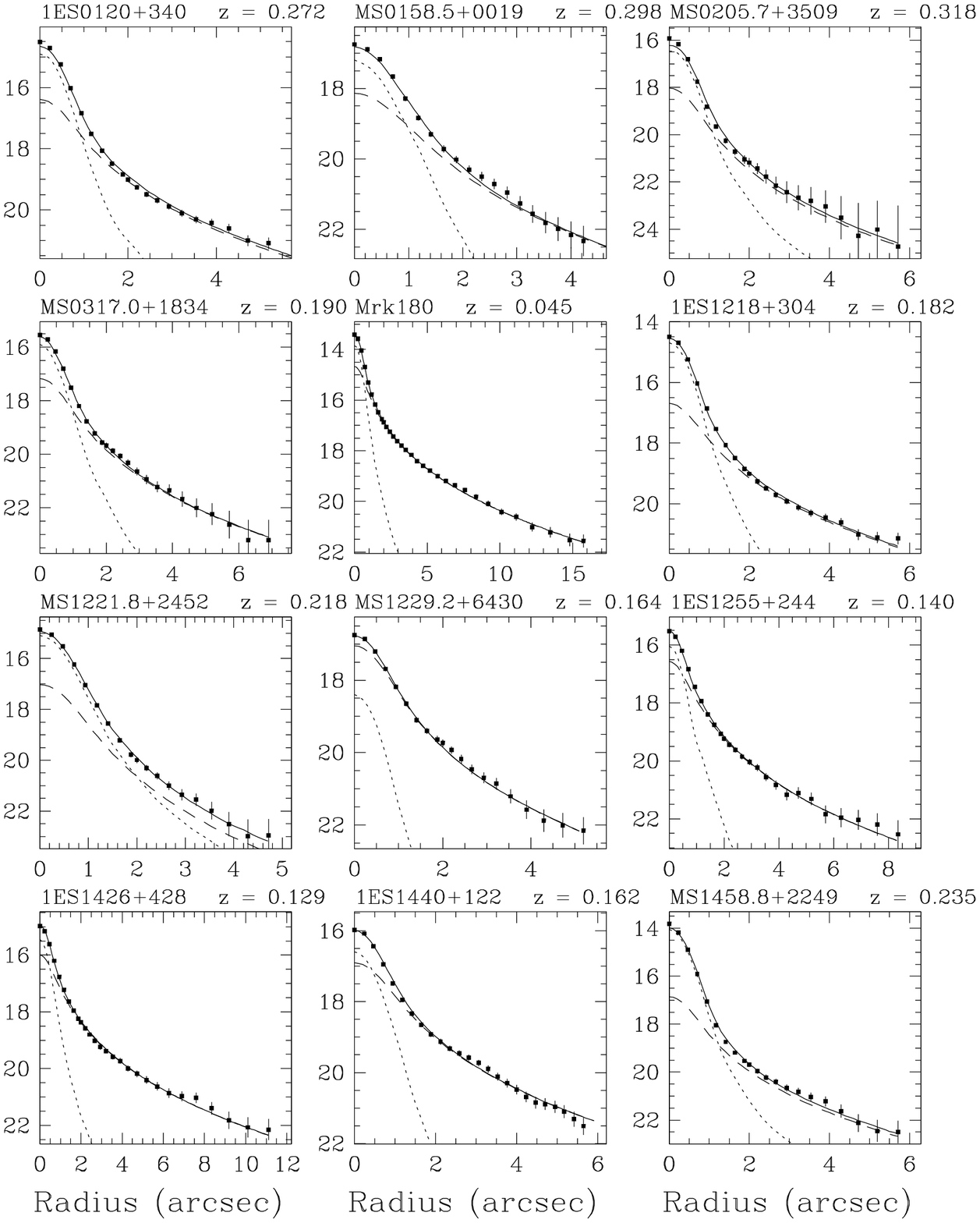}
\caption{
The observed $H$-band azimuthally averaged radial surface brightness profile 
(solid points with error bars) for each BL Lac object, overlaid with the 
scaled PSF model (dotted line), the de Vaucouleurs $r^{1/4}$ model convolved 
with the PSF (dashed line), and the fitted PSF + host model profile 
(solid line). The y-axis is plotted in mag arcsec$^{-2}$.
}
\label{blfig2}
\end{figure*}
\addtocounter{figure}{-1}%

\begin{figure*}
\centering
\includegraphics[width=15cm]{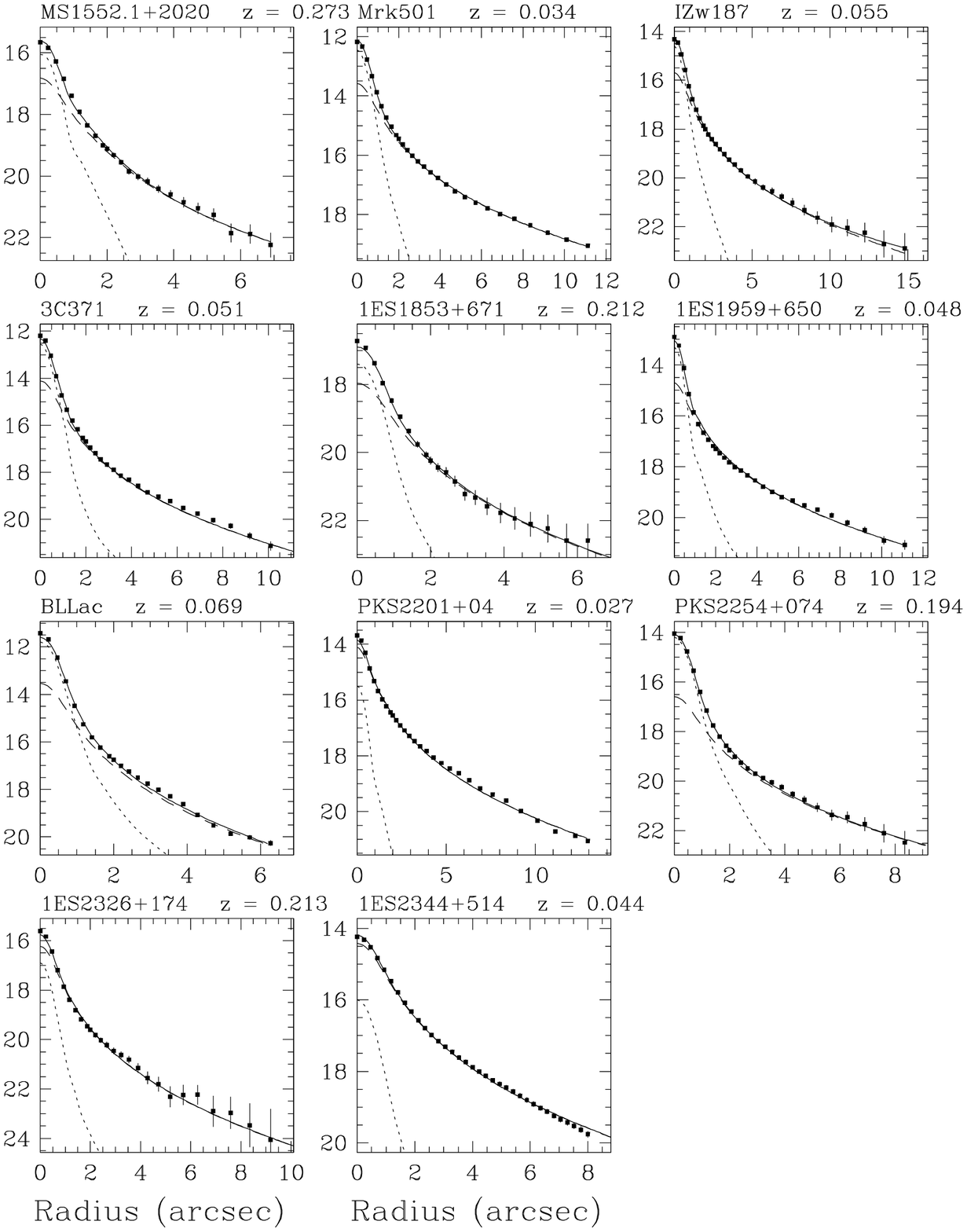}
\caption{Continued.}
\label{blfig2}
\end{figure*}

\begin{figure}
\centering
\includegraphics[width=15cm]{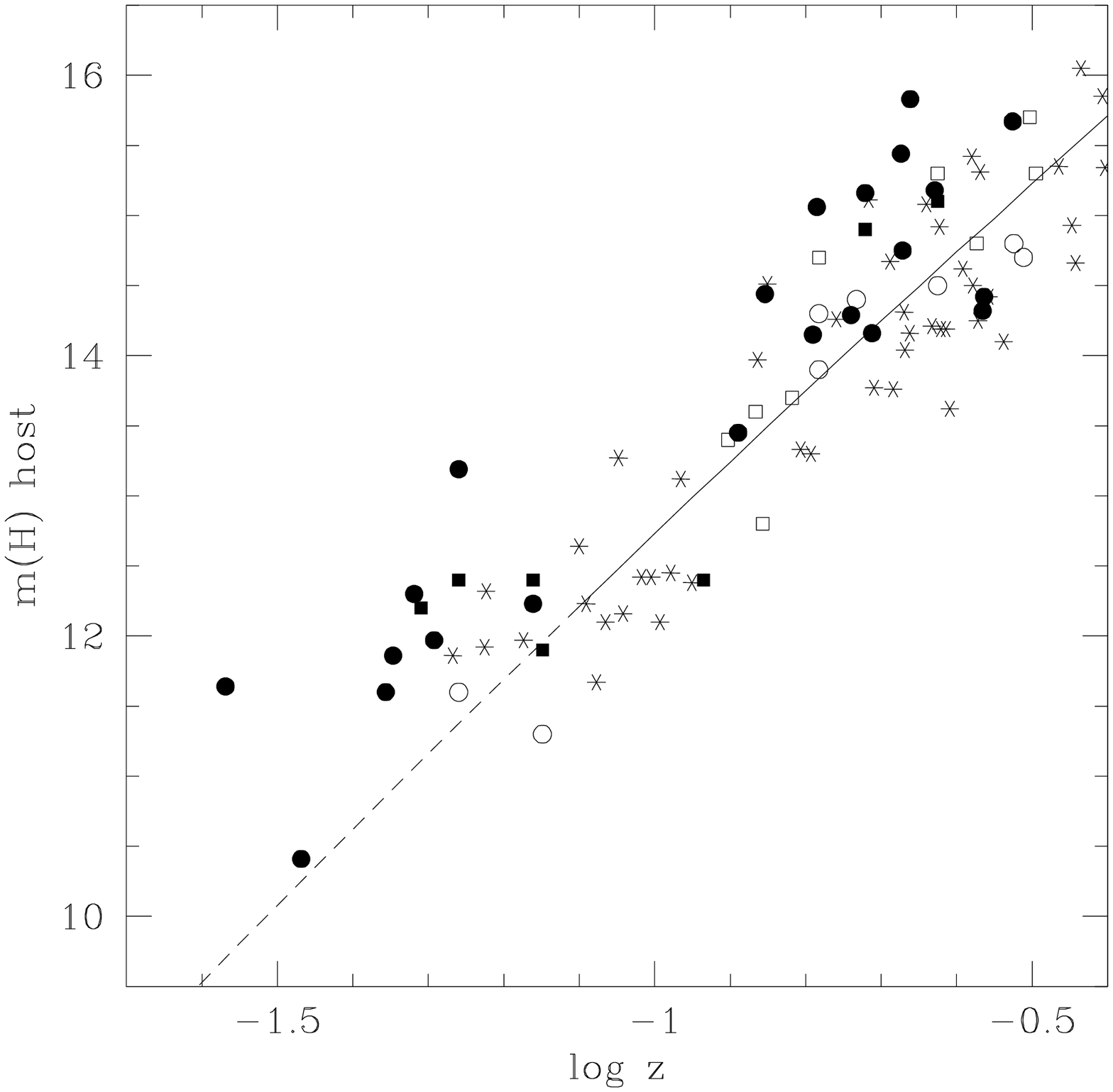}
\caption{Plot of the apparent H magnitude of the host galaxies vs. redshift 
(Hubble diagram). The BL Lacs are marked as filled circles (this work), 
filled squares (K98), open squares (S00) and open circles (C03). The solid 
line is the $K$-z relation for RGs derived by Willott et al. (2003), 
converted to the $H$-band assuming $H$--$K$ = 0.2. The dashed line is the 
extrapolation of the relation towards lower redshift. The RGs are shown 
as asterisks.}
\label{mhz}
\end{figure}

\begin{figure}
\centering
\includegraphics[width=15cm]{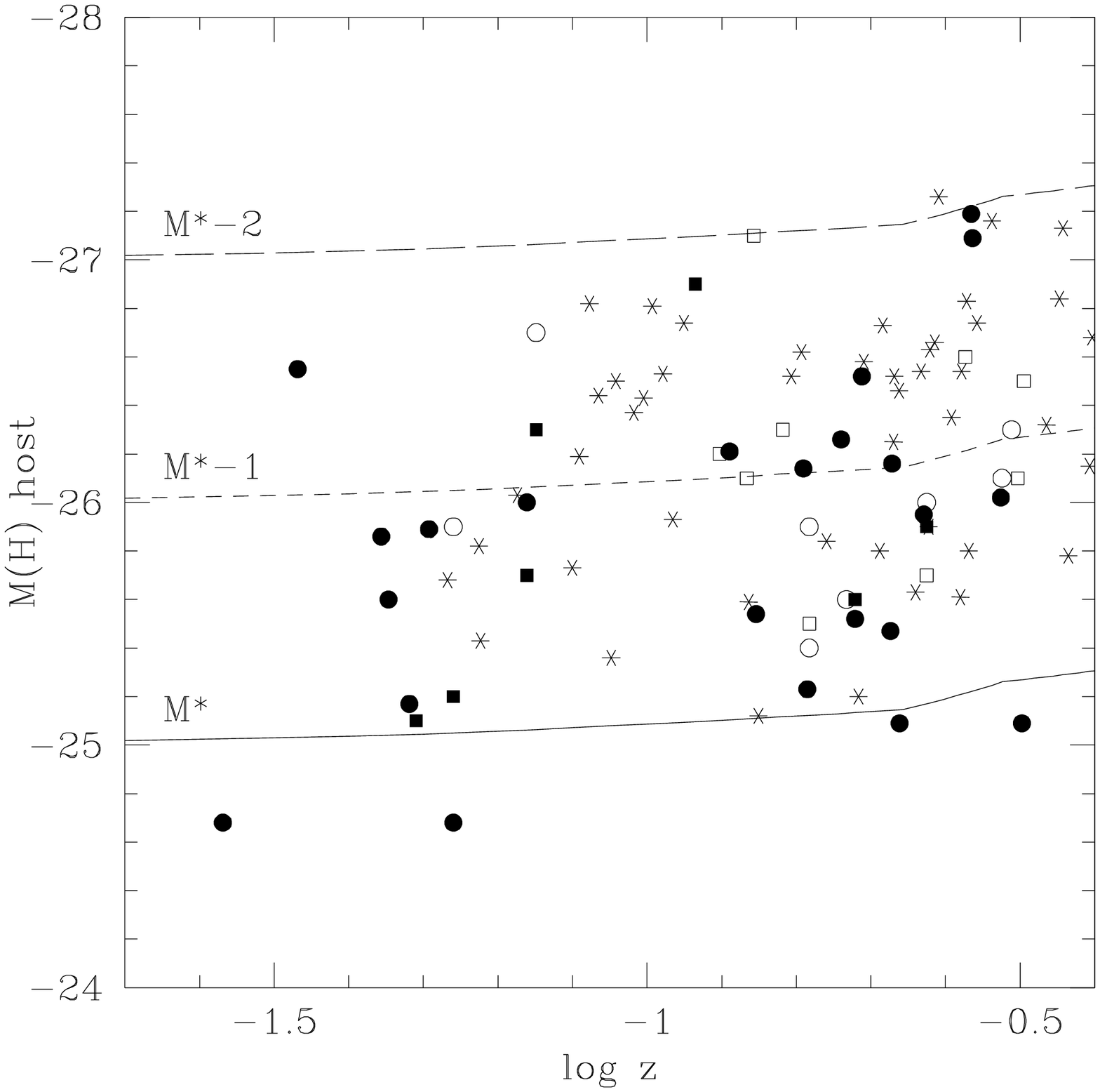}
\caption{
Plot of the absolute H band magnitude of the host galaxies 
vs. redshift. For symbols, see Fig.~\ref{mhz}. The solid, short-dashed and 
long-dashed lines are the luminosities of L* (M(H)$\sim$-25.0 at low redshift; 
Mobasher et al. 1993), L*-1 and L*-2 galaxies, respectively, following the 
passive evolution model of Bressan, Granato \& Silva (1998). 
}
\label{Mhz}
\end{figure}

\begin{figure}
\centering
\includegraphics[width=15cm]{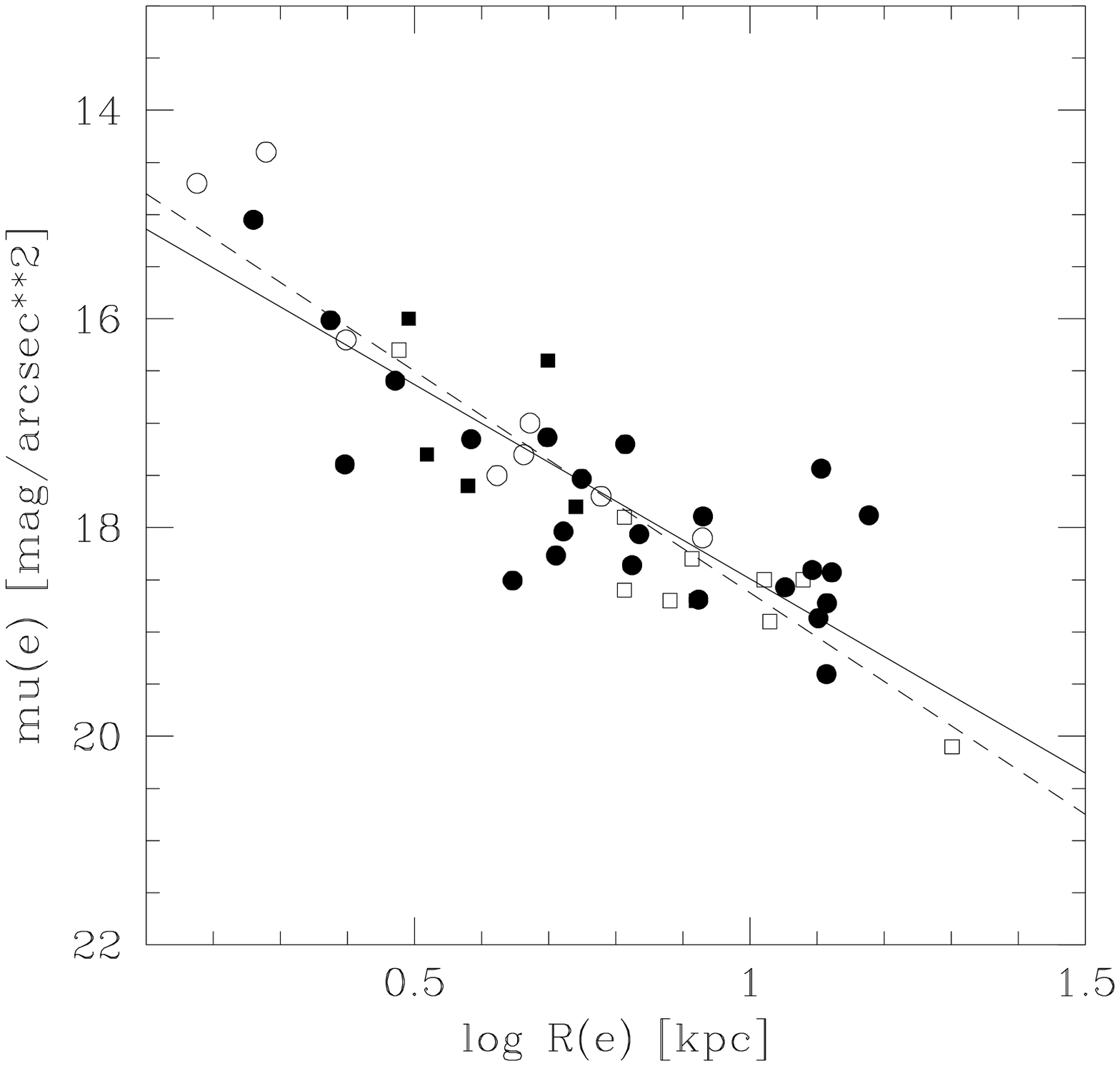}
\caption{The $H$-band $\mu_e$ - r$_e$ relation for the full sample of BL Lac 
host galaxies.  For symbols, see Fig.~\ref{mhz}. 
The solid and dashed lines are the linear least-square best-fit relations for 
the BL Lacs and for normal inactive ellipticals (Pahre et al. 1995), 
respectively.
}
\label{muere}
\end{figure}

\begin{figure}
\centering
\includegraphics[width=15cm]{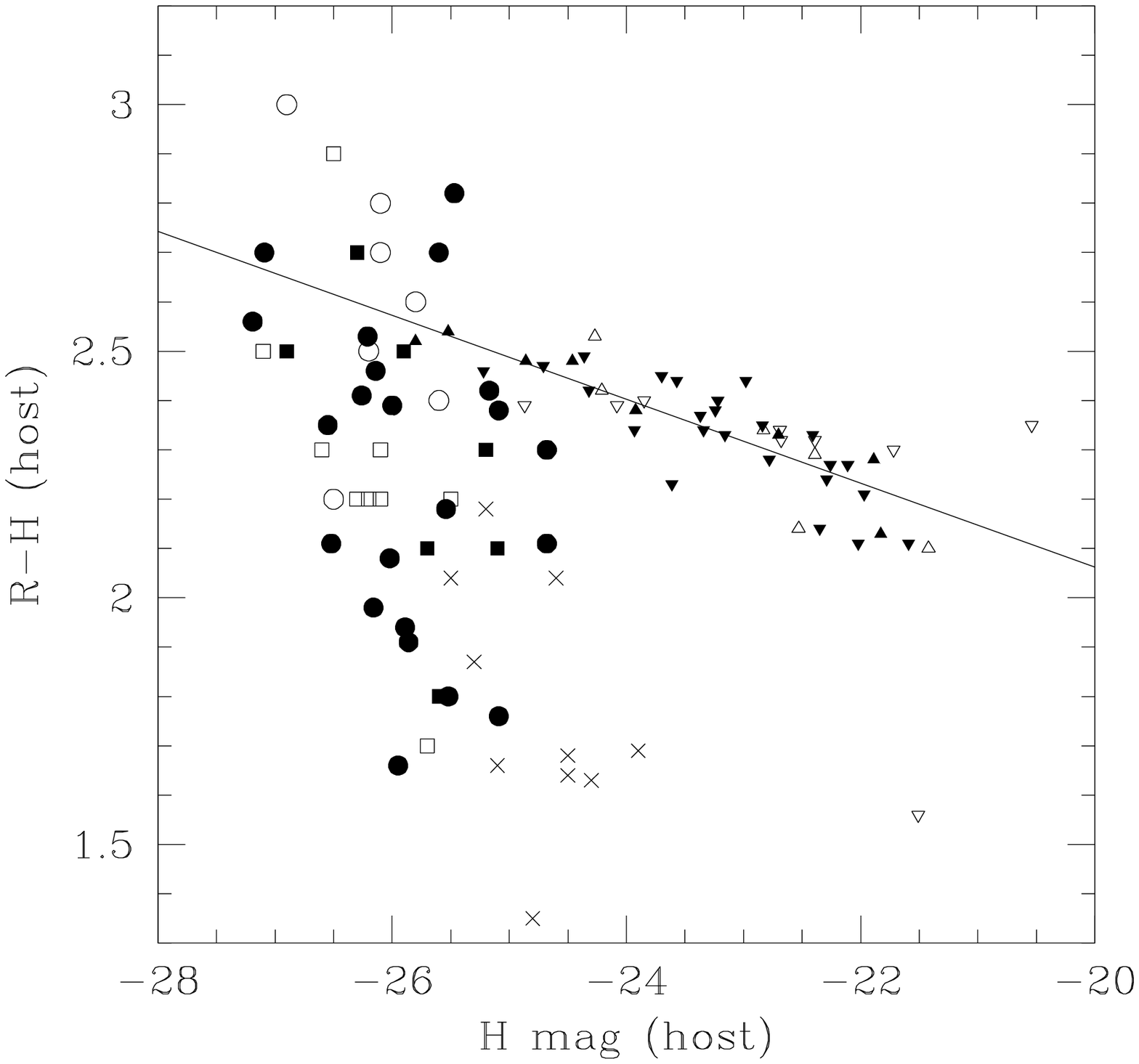}
\caption{
The $R$--$H$ vs. H colour-magnitude diagram for the BL Lac host galaxies. 
For symbols, see Fig.~\ref{mhz}. The other symbols indicate low redshift 
(z $<$ 0.2) elliptical quasar hosts (crosses; from Jahnke et al. 2003) 
and elliptical galaxies (filled triangles) and S0 galaxies 
(open triangles) in the Virgo cluster, and elliptical galaxies 
(filled inverted triangles) and S0 galaxies (open inverted triangles) 
in the Coma cluster (from Bower et al. 1992a,b). 
The solid line shows the best--fit regression line for Virgo and Coma clusters 
(Bower et al. 1992b). The $V$--$K$ vs. V diagram of Bower et al. (1992b) 
has been transformed into $R$--$H$ vs. H assuming $V$--$R$ = 0.6, 
$H$--$K$ = 0.2 (see text), and distance moduli for Virgo and Coma clusters 
m--M = 31.0 and 34.6, respectively (Bower et al. 1992b).
}
\label{hrh}
\end{figure}

\begin{figure*}
\centering
\includegraphics[width=15cm]{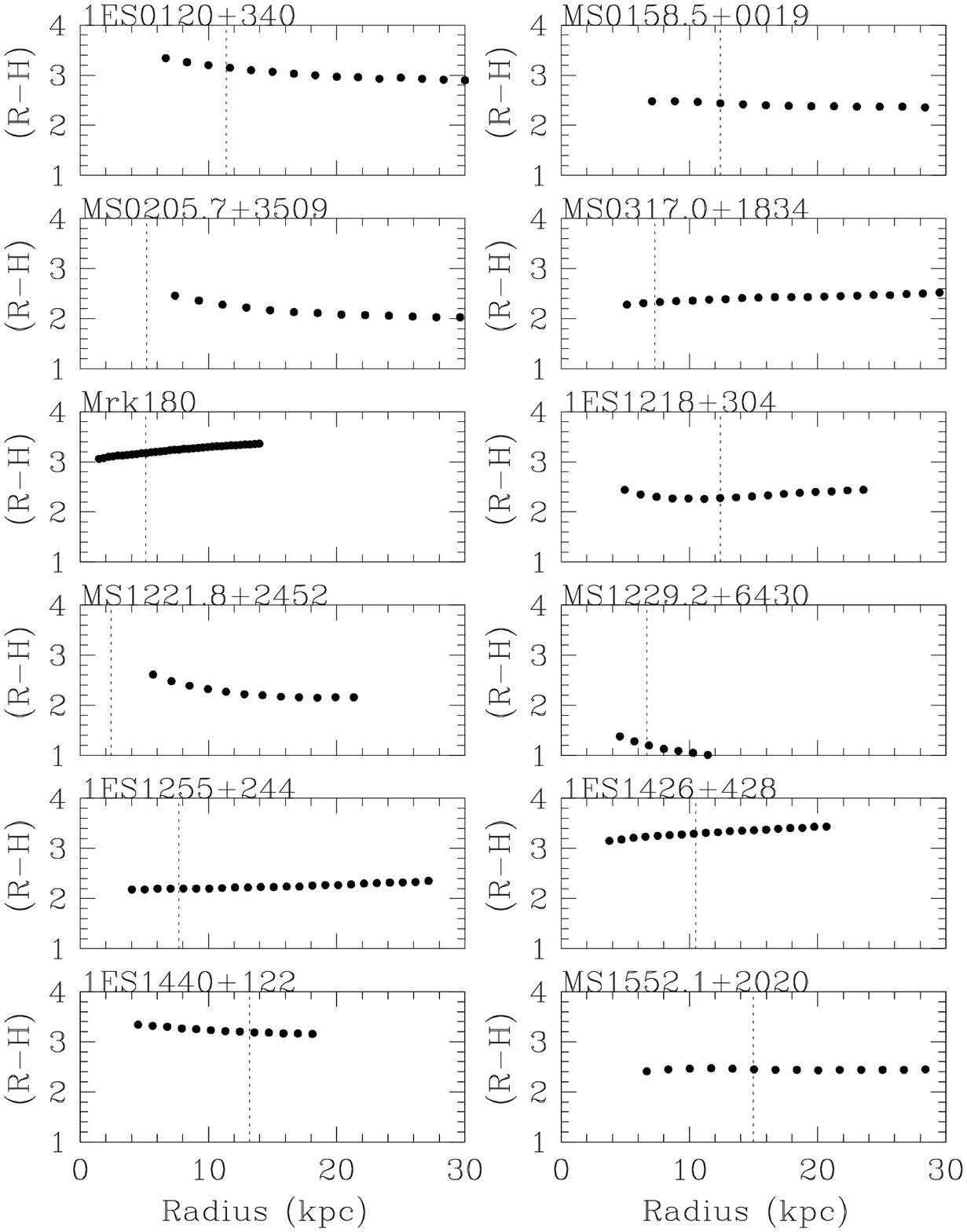}
\caption{
The host galaxy $R$--$H$ radial colour profile for all the BL Lacs in 
the sample, calculated using the $R$-band data from 
Falomo \& Kotilainen (1999) and Falomo \& Kotilainen (in prep.). 
The dashed vertical lines refer to the effective radius derived from 
the fit of the $R$-band data.
The profiles 
indicate that the host galaxies are predominantly bluer further away from 
the nuclei, but there are marked exceptions with bluer inner regions, 
e.g. MS 0317, Mrk 180 and 1ES 1853. 
}
\label{blfig3}
\end{figure*}
\addtocounter{figure}{-1}%

\begin{figure*}
\centering
\includegraphics[width=15cm]{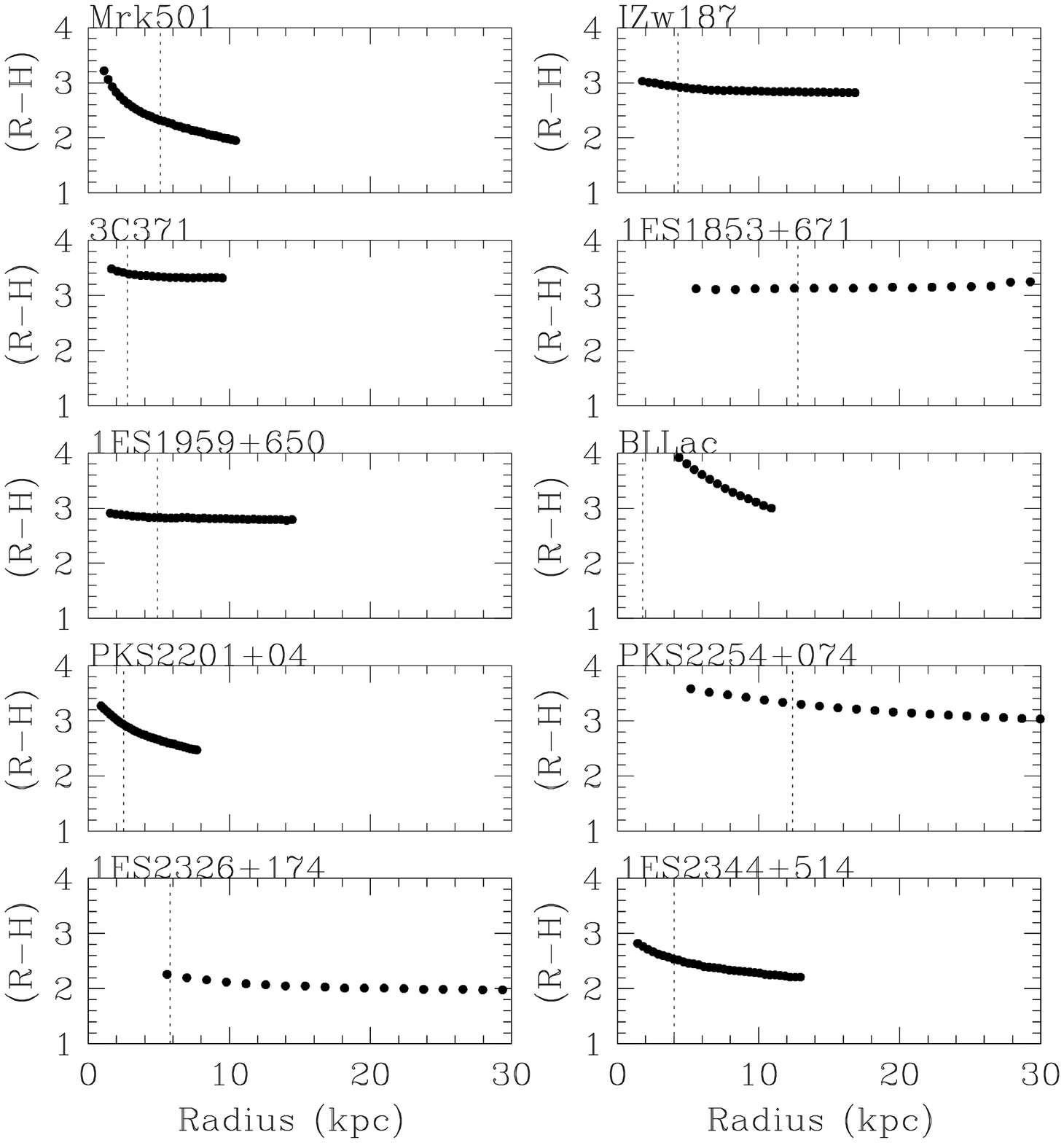}
\caption{Continued.}
\label{blfig3}
\end{figure*}

\begin{figure}
\centering
\includegraphics[width=15cm]{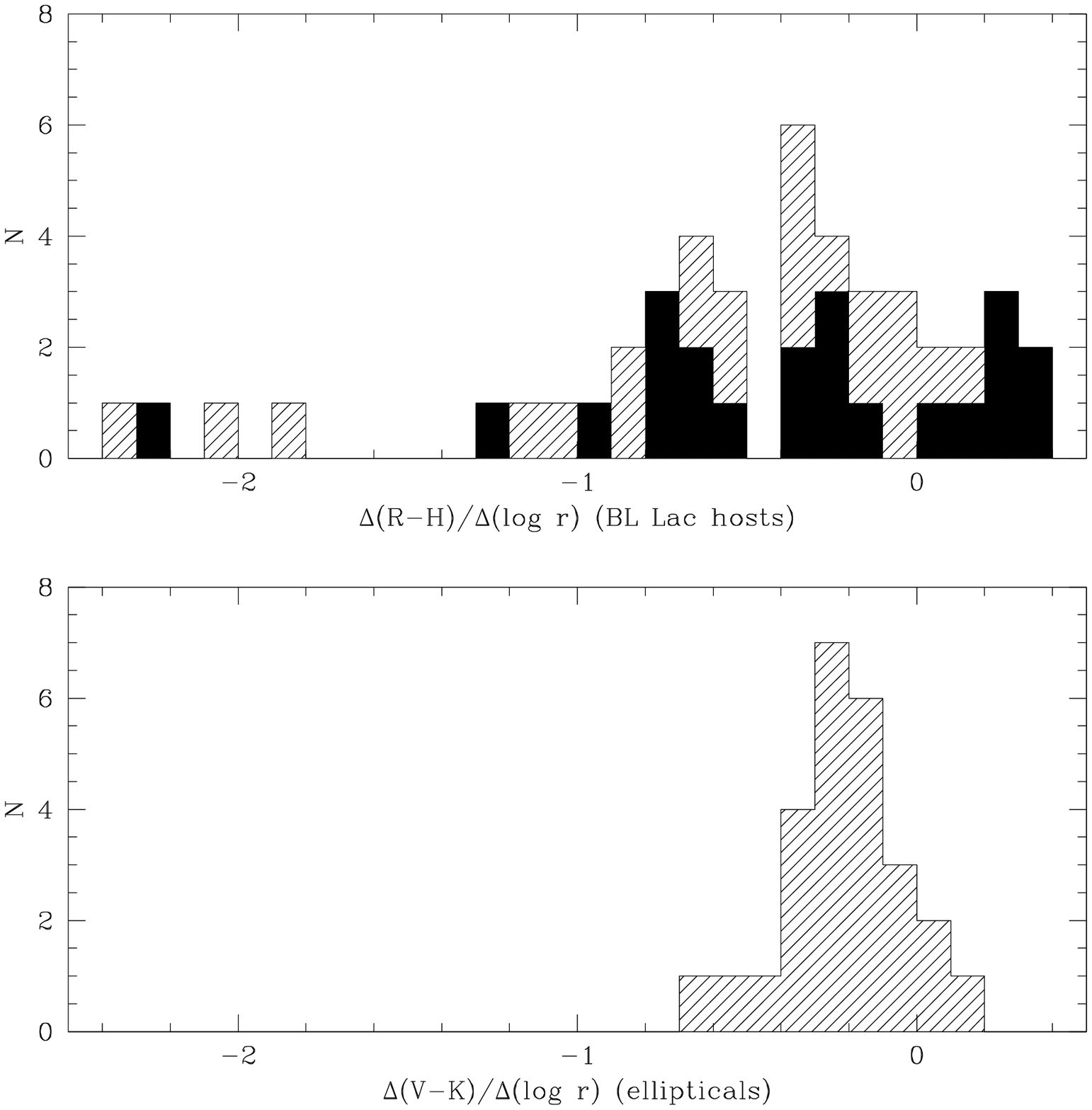}
\caption{
{\bf Upper panel:} Histogram of the host galaxy $R$-$H$ colour gradient 
for all the BL Lac samples. The solid histogram shows the BL Lacs in 
this work. 
{\bf Lower panel:} Histogram of the host galaxy $V$-$K$ colour gradient for 
normal ellipticals (from Peletier et al. 1990 and Schombert et al. 1993).
}
\label{colgradhist}
\end{figure}

\begin{table}
\caption[]{The sample and journal of observations.$^{\mathrm{a}}$}
\label{sample}
\begin{flushleft}
\begin{tabular}{lllllrll}
\hline
\noalign{\smallskip}
Name & L/H & z & V & M(B) & Date & T(exp) & FWHM \\
& & & & & & min & arcsec\\
(1) & (2) & (3) & (4) & (5) & (6) & (7) & (8)\\
\noalign{\smallskip}
\hline
\noalign{\smallskip}
1ES 0120+340   & H & 0.272  & 15.2 & -26.0 & 20/07/2002    & 32 & 0.75\\
MS 0158.5+0019 & H & 0.298  & 18.0 & -23.4 & 17-18/07/2002 & 44 & 0.85\\
MS 0205.7+3509 & H & 0.318: & 19.2 & -22.3 & 20/07/2002    & 48 & 0.65\\
MS 0317.0+1834 & H & 0.190  & 18.1 & -22.2 & 18/07/2002    & 32 & 0.80\\
Mrk 180        & H & 0.045  & 14.5 & -22.1 & 17/07/2002    & 20 & 0.60\\
1ES 1218+304   & H & 0.182  & 16.5 & -23.0 & 18/07/2002    & 32 & 0.70\\
MS 1221.8+2452 & H & 0.218: & 17.4 & -22.7 & 19/07/2002    & 32 & 0.85\\
MS 1229.2+6430 & H & 0.164  & 16.9 & -23.1 & 17/07/2002    & 40 & 0.60\\
1ES 1255+244   & H & 0.140  & 15.4 & -24.3 & 20/07/2002    & 27 & 0.70\\
1ES 1426+428   & H & 0.129  & 16.5 & -22.5 & 17/07/2002    & 30 & 0.70\\
1ES 1440+122   & H & 0.162  & 17.1 & -22.9 & 18/07/2002    & 32 & 0.90\\
MS 1458.8+2249 & H & 0.235  & 16.8 & -24.0 & 20/07/2002    & 32 & 0.60\\
MS 1552.1+2020 & H & 0.273  & 17.7 & -23.0 & 18/07/2002    & 40 & 0.80\\
Mrk 501        & H & 0.034  & 13.8 & -22.4 & 17/07/2002    & 10 & 0.70\\
I Zw 187       & H & 0.055  & 16.0 & -21.1 & 17/07/2002    & 30 & 0.60\\
3C 371         & L & 0.051  & 14.2 & -22.7 & 18/07/2002    & 23 & 0.75\\
1ES 1853+671   & H & 0.212  & 16.4 & -24.2 & 17/07/2002    & 30 & 0.85\\
1ES 1959+650   & H & 0.048: & 12.8 & -24.5 & 20/07/2002    & 16 & 0.55\\
BL Lac         & L & 0.069  & 14.7 & -22.4 & 17/07/2002    & 28 & 0.65\\
PKS 2201+04    & L & 0.027  & 15.2 &       & 19/07/2002    & 32 & 0.65\\
PKS 2254+074   & L & 0.194  & 16.4 & -23.3 & 19/07/2002    & 32 & 0.70\\
1ES 2326+174   & H & 0.213  & 16.8 & -23.8 & 17/07/2002    & 40 & 0.70\\
1ES 2344+514   & H & 0.044  & 15.5 & -21.6 & 18/07/2002    & 24 & 0.75\\
\noalign{\smallskip}
\hline
\end{tabular}
\end{flushleft}
\begin{list}{}{}
\item[$^{\mathrm{a}}$] 
Column (1) gives the name of the object; (2) high-frequency (H) or 
low-frequency (L) peaked object; (3) the redshift; 
(4) V-band apparent magnitude; (5) B-band absolute magnitude; 
(6) the date of observation; (7) total exposure time; and (8) seeing FWHM
\end{list}
\end{table}

\begin{table}
\caption[]{Properties of the host galaxies.$^{\mathrm{a}}$}
\label{hostprop}
\begin{flushleft}
\begin{tabular}{lllllllrlll}
\hline
\noalign{\smallskip}
 & z & A$_H$$^{\mathrm{b}}$ & m$_H$(nuc) & m$_H$(gal) & $\mu_e$$^{\mathrm{c}}$ & r$_e$ & R$_e$ & M$_H$(nuc) & M$_H$(gal) & N/G\\
 &  & mag &  &  &  & arcsec & kpc &  &  & \\
(1) & (2) & (3) & (4) & (5) & (6) & (7) & (8) & (9) & (10) & (11)\\
\noalign{\smallskip}
\hline
\noalign{\smallskip}
1ES 0120+340   & 0.272 & 0.06 & 14.57 & 14.32 & 17.44 & 2.30 & 12.8 & -26.77 & -27.19 & 0.68\\
MS 0158.5+0019 & 0.298 & 0.03 & 16.54 & 15.67 & 18.41 & 2.10 & 12.4 & -25.02 & -26.02 & 0.40\\ 
MS 0205.7+3509 & 0.318 & 0.07 & 16.15 & 16.76 & 17.53 & 0.90 & 5.6  & -25.57 & -25.09 & 1.6\\
MS 0317.0+1834 & 0.190 & 0.11 & 15.58 & 15.16 & 18.06 & 1.60 & 6.8  & -24.90 & -25.52 & 0.56\\
Mrk 180        & 0.045 & 0.02 & 13.71 & 11.86 & 18.04 & 4.30 & 5.3  & -23.50 & -25.60 & 0.14 \\
1ES 1218+304   & 0.182 & 0.02 & 14.51 & 14.29 & 18.72 & 3.15 & 13.0 & -25.87 & -26.26 & 0.70 \\
MS 1221.8+2452 & 0.218 & 0.02 & 14.69 & 15.83 & 16.01 & 0.50 & 2.4  & -26.12 & -25.09 & 2.6\\
MS 1229.2+6430 & 0.164 & 0.02 & 18.32 & 15.06 & 18.36 & 1.75 & 6.7  & -21.82 & -25.23 & 0.04\\
1ES 1255+244   & 0.140 & 0.01 & 15.91 & 14.44 & 18.69 & 2.50 & 8.4  & -23.86 & -25.54 & 0.21\\
1ES 1426+428   & 0.129 & 0.02 & 15.37 & 13.45 & 18.57 & 3.60 & 11.3 & -24.21 & -26.21 & 0.16\\
1ES 1440+122   & 0.162 & 0.02 & 16.23 & 14.15 & 18.87 & 3.35 & 12.6 & -23.88 & -26.14 & 0.12\\
MS 1458.8+2249 & 0.235 & 0.04 & 14.21 & 15.18 & 17.89 & 1.70 & 8.5  & -26.78 & -25.95 & 2.1\\
MS 1552.1+2020 & 0.273 & 0.04 & 15.83 & 14.42 & 17.88 & 2.70 & 15.0 & -25.52 & -27.09 & 0.24\\
Mrk 501        & 0.034 & 0.02 & 12.22 & 10.41 & 17.13 & 5.30 & 5.0  & -24.36 & -26.55 & 0.13\\
I Zw 187       & 0.055 & 0.03 & 14.44 & 13.19 & 18.51 & 3.00 & 4.4  & -23.21 & -24.68 & 0.26\\
3C 371 	       & 0.051 & 0.05 & 12.53 & 11.97 & 16.59 & 2.15 & 3.0  & -24.95 & -25.89 & 0.42\\
1ES 1853+671   & 0.212 & 0.06 & 16.87 & 15.44 & 19.40 & 2.80 & 13.0 & -23.87 & -25.47 & 0.23\\
1ES 1959+650   & 0.048 & 0.11 & 13.69 & 12.30 & 18.27 & 3.95 & 5.1  & -23.66 & -25.17 & 0.25\\
BL Lac         & 0.069 & 0.21 & 11.90 & 12.23 & 15.05 & 1.00 & 1.8  & -26.26 & -26.00 & 1.3\\
PKS 2201+04    & 0.027 & 0.06 & 15.50 & 11.64 & 17.39 & 3.30 & 2.5  & -20.58 & -24.68 & 0.02\\
PKS 2254+074   & 0.194 & 0.06 & 13.96 & 14.16 & 18.43 & 3.05 & 13.2 & -26.57 & -26.52 & 1.0\\
1ES 2326+174   & 0.213 & 0.04 & 17.12 & 14.75 & 17.20 & 1.40 & 6.5  & -23.63 & -26.16 & 0.10\\
1ES 2344+514   & 0.044 & 0.17 & 15.69 & 11.60 & 17.15 & 3.20 & 3.8  & -21.46 & -25.86 & 0.02\\
\noalign{\smallskip}
\hline
\end{tabular}
\end{flushleft}
\begin{list}{}{}
\item[$^{\mathrm{a}}$] 
Column (1) and (2) give the name and redshift of the object; 
(3) the interstellar extinction correction; 
(4) and (5) the apparent nuclear and host galaxy magnitude; 
(6) the surface brightness $\mu$(e); (7) and (8) the bulge scalelength in 
arcsec and kpc; (9) and (10) the absolute nuclear and host galaxy magnitude; 
and (11) the nucleus/galaxy luminosity ratio. 
\item[$^{\mathrm{b}}$] 
The interstellar extinction corrections were computed using R-band extinction 
coefficient from Urry et al. (2000) and A$_H$/A$_R$ = 0.234 
(Cardelli et al. 1989).
\item[$^{\mathrm{c}}$] 
Corrected for galactic extinction and cosmological dimming.
\end{list}
\end{table}

\begin{table}
\caption[]{Comparison with previous studies for common BL Lacs.}
\label{compbl}
\begin{flushleft}
\begin{tabular}{llllll}
\hline
\noalign{\smallskip}
Object & R$_e$ & M$_H$(gal) & $R$-$H$ & $\Delta$($R$-$H$)/$\Delta$(log r) & Ref.\\
(1) & (2) & (3) & (4) & (5) & (6) \\
\noalign{\smallskip}
\hline
\noalign{\smallskip}
MS 0158.5+0019 & 12 & -26.0 & 2.1 & -0.22 & this work\\
               & 4.7 & -26.3 & 3.3 & -2.0 & C03\\
PKS 2254+074   & 13 & -26.5 & 2.1 & -0.75 & this work\\
               & 32 & -25.6 & 1.8 & 0.10 & K98\\
MS 0317.0+1834 & 6.8 & -25.5 & & & this work\\
               & 14 & -25.2 & & & Wright et al. (1998)\\
\noalign{\smallskip}
\hline
\end{tabular}
\end{flushleft}
\end{table}

\begin{table}
\caption[]{Comparison of the average NIR host galaxy properties.$^{\mathrm{a}}$}
\label{avgprop}
\begin{flushleft}
\begin{tabular}{llrllll}
\hline
\noalign{\smallskip}
Sample & filter & N & $<z>$ & $<M_H(nuc)>$ & $<M_H(host)>$ & $<R(e)>$\\
(1) & (2) & (3) & (4) & (5) & (6) & (7) \\
\noalign{\smallskip}
\hline
\noalign{\smallskip}
BL (this work) & H & 23 & 0.155$\pm$0.091 & -24.5$\pm$1.6 & -25.8$\pm$0.7 & 7.8$\pm$4.1\\
BL C03 & K   & 8  & 0.186$\pm$0.088 & -25.6$\pm$0.8 & -26.0$\pm$0.4 & 4.2$\pm$2.3\\
BL S00 & H   & 9  & 0.206$\pm$0.074 & -25.0$\pm$1.6 & -26.2$\pm$0.4 & 10$\pm$5\\
BL K98 & H   & 7  & 0.112$\pm$0.068 & -25.7$\pm$1.7 & -25.8$\pm$0.5 &  4.8$\pm$1.9\\
BL all & H/K & 42 & 0.164$\pm$0.084 & -25.0$\pm$1.5 & -25.9$\pm$0.6 & 7.2$\pm$3.6\\
	     &	 &   &                 &               &             & \\
L* Mobasher et al. (1993) & K & 136 & 0.077$\pm$0.030 & & -25.0$\pm$0.2 & \\
	     &	 &   &                 &               &             & \\
BCM Thuan \& Puschell (1989) & H & 84 & 0.074$\pm$0.026 & & -26.3$\pm$0.3 & \\
BCM Aragon-Salamanca et al. (1998) & K & 25 & 0.449$\pm$0.266 & & -27.0$\pm$0.3 & \\
	&	&   &       &         &             & \\
RG FR II Taylor et al. (1996) & K & 12 & 0.214$\pm$0.049 & -25.1$\pm$0.7 & -26.1$\pm$0.8 & 26$\pm$16\\
RG Willott et al. (2003) z $<$ 0.3 & K & 42 & 0.170$\pm$0.075 & & -26.2$\pm$0.5 & \\
\noalign{\smallskip}
\hline
\end{tabular}
\end{flushleft}
\begin{list}{}{}
\item[$^{\mathrm{a}}$] 
Column (1) gives the sample; (2) the filter; (3) the number of objects in 
the sample; (4) the average redshift of the sample; (5) and (6) the average 
H band nuclear and host galaxy absolute magnitude of the sample; 
and (7) the average H-band bulge scale length of the sample.
\end{list}
\end{table}

\begin{table}
\caption[]{Optical--NIR colours of the host galaxies.$^{\mathrm{a}}$}
\label{colour}
\begin{flushleft}
\begin{tabular}{llllllr}
\hline
\noalign{\smallskip}
Name & z & M(H) & M(R) & Ref.$^{\mathrm{b}}$ & $R$-$H$ & $\Delta$($R$-$H$)/$\Delta$(log r) \\
(1) & (2) & (3) & (4) & (5) & (6) & (7) \\
\noalign{\smallskip}
\hline
\noalign{\smallskip}
1ES 0120+340   & 0.272 & -27.2 & -24.6 & 1 & 2.6 & -0.66 \\
MS 0158.5+0019 & 0.298 & -26.0 & -23.9 & 1 & 2.1 & -0.22 \\
MS 0205.7+3509 & 0.318 & -25.1 & -23.3 & 1 & 1.8 & -0.52 \\
MS 0317.0+1834 & 0.190 & -25.5 & -23.7 & 1 & 1.8 & 0.29 \\
Mrk 180        & 0.045 & -25.6 & -22.9 & 2 & 2.7 & 0.33 \\
1ES 1218+304   & 0.182 & -26.3 & -23.8 & 1 & 2.5 & 0.11 \\
MS 1221.8+2452 & 0.218 & -25.1 & -22.7 & 1 & 2.4 & -0.78 \\
MS 1229.2+6430 & 0.164 & -25.2 & -24.3 & 1 & (0.9) & -0.77 \\
1ES 1255+244   & 0.140 & -25.5 & -23.4 & 1 & 2.1 & 0.20 \\
1ES 1426+428   & 0.129 & -26.2 & -23.7 & 2 & 2.5 & 0.38 \\
1ES 1440+122   & 0.162 & -26.1 & -23.7 & 2 & 2.4 & -0.32 \\
MS 1458.8+2249 & 0.235 & -26.0 & -24.3 & 1 & 1.7 & \\
MS 1552.1+2020 & 0.273 & -27.1 & -24.4 & 1 & 2.7 & 0.04 \\
Mrk 501        & 0.034 & -26.6 & -24.2 & 2 & 2.4 & -1.23 \\
I Zw 187       & 0.055 & -24.7 & -22.4 & 2 & 2.3 & -0.21 \\
3C 371 	       & 0.051 & -25.9 & -24.0 & 2 & 1.9 & -0.20 \\
1ES 1853+671   & 0.212 & -25.5 & -22.6 & 1 & 2.9 & 0.21 \\
1ES 1959+650   & 0.048 & -25.2 & -22.8 & 1 & 2.4 & -0.12 \\
BL Lac         & 0.069 & -26.0 & -23.6 & 2 & 2.4 & -2.25 \\
PKS 2201+04    & 0.027 & -24.7 & -22.6 & 2 & 2.1 & -0.90 \\
PKS 2254+074   & 0.194 & -26.5 & -24.4 & 2 & 2.1 & -0.75 \\
1ES 2326+174   & 0.213 & -26.2 & -24.2 & 1 & 2.0 & -0.34 \\
1ES 2344+514   & 0.044 & -25.9 & -24.0 & 1 & 1.9 & -0.65 \\
\noalign{\smallskip}
\hline
\end{tabular}
\end{flushleft}
\begin{list}{}{}
\item[$^{\mathrm{a}}$] 
Columns (1) and (2) give the name and redshift of the BL Lac; 
(3) and (4) the absolute magnitude of the host galaxy in the $H$-band 
(this work) and $R$-band (literature values), respectively; 
(5) the references for column (4); (6) the $R$--$H$ colour of the host galaxy 
computed from columns (3) and (4); and (7) the $R$--$H$ colour gradient of 
the host. 
\item[$^{\mathrm{b}}$] 
References: 1 = Falomo \& Kotilainen (1999); 2 = Falomo \& Kotilainen (in prep.).
\end{list}
\end{table}

\end{document}